\documentclass[aps,prb,twocolumn,showpacs,floatfix,superscriptaddress]{revtex4}
\usepackage{dcolumn}
\usepackage{bm}
\usepackage{amsmath,amssymb,graphicx}

\begin{document}
\preprint{Submitted to Phys. Rev. B}

\title{Midinfrared Third Harmonic Generation from Macroscopically Aligned \\Ultralong Single-Wall Carbon Nanotubes}

\author{D.~T.~Morris}
\affiliation{Department of Electrical and Computer Engineering, Rice University, Houston, Texas 77005, USA}

\author{C.~L.~Pint}
\affiliation{Department of Chemistry, Rice University, Houston, Texas 77005, USA}
\affiliation{Department of Mechanical Engineering, Vanderbilt University, Nashville, Tennessee 37240, USA}

\author{R.~S.~Arvidson}
\thanks{Present address: MARUM, University of Bremen, Germany}
\affiliation{Department of Earth Science, Rice University, Houston, Texas 77005, USA}

\author{A.~L\"uttge}
\thanks{Present address: MARUM, University of Bremen, Germany}
\affiliation{Department of Earth Science, Rice University, Houston, Texas 77005, USA}
\affiliation{Department of Chemistry, Rice University, Houston, Texas 77005, USA}

\author{R.~H.~Hauge}
\affiliation{Department of Chemistry, Rice University, Houston, Texas 77005, USA}

\author{A.~A.~Belyanin}
\affiliation{Department of Physics and Astronomy, Texas A\&M University, College Station, Texas 77843, USA}

\author{G.~L.~Woods}
\affiliation{Department of Electrical and Computer Engineering, Rice University, Houston, Texas 77005, USA}

\author{J.~Kono}
\email[]{kono@rice.edu}
\thanks{corresponding author.}
\affiliation{Department of Electrical and Computer Engineering, Rice University, Houston, Texas 77005, USA}
\affiliation{Department of Physics and Astronomy, Rice University, Houston, Texas 77005, USA}

\date{\today}
\begin{abstract}
We report the observation of strong third harmonic generation from a macroscopic array of aligned ultralong single-wall carbon nanotubes (SWCNTs) with intense midinfrared radiation. Through power-dependent experiments, we determined the absolute value of the third-order nonlinear optical susceptibility, $\chi^{(3)}$, of our SWCNT film to be 5.53 $\times$ 10$^{-12}$~esu, three orders of magnitude larger than that of the fused silica reference we used.  Taking account of the filling factor of 8.75\% for our SWCNT film, we estimate a $\chi^{(3)}$ of 6.32 $\times$ 10$^{-11}$~esu for a fully dense film.  Furthermore, through polarization-dependent experiments, we extracted all the nonzero elements of the $\chi^{(3)}$ tensor, determining the magnitude of the weaker tensor elements to be $\sim$1/6 of that of the dominant $\chi^{(3)}_{zzzz}$ component. 
\end{abstract}

\maketitle


Carbon nanomaterials, i.e., carbon nanotubes and graphene, attract much attention both from fundamental and applied viewpoints.  These novel low-dimensional systems possess unique band structure and extraordinary properties that are promising for a variety of applications.\cite{AvourisetAl08NP,BonaccorsoetAlNP10}  Single-wall carbon nanotubes (SWCNTs), in particular, are ideal one-dimensional systems for basic optical studies as well as for multi-wavelength photonic devices due to their diameter-dependent, direct band gaps.\cite{NanotetAl12AM}  Optical properties of SWCNTs have been extensively studied during the last decade, and much basic knowledge has been accumulated on how light emission, scattering, and absorption occur in the realm of linear optics.\cite{DresselhausetAl07ARPC,AvourisetAl08NP,NanotetAl12AM}  However, {\em nonlinear} optical properties of carbon nanomaterials remain largely unexplored although a number of interesting predictions exist.\cite{AlonetAl00PRL,GumbsetAl03EPL,Alon03PRB,DumitricaetAl04PRL,Mikhailov07EPL,OkaAoki09PRB,CalvoetAl11APL,YaoBelyanin12PRL}

Theoretical calculations predict large non-resonant third-order nonlinear optical susceptibilities, $\chi^{(3)} \approx 10^{-8}$ to $10^{-6}$\,esu, for SWCNTs, which vary rapidly with the tube diameter.\cite{XieJiang97APL,MargulisSizikova98Physica,JieetAl99PRB,MargulisetAl99PLA,MargulisGaiduk01JOA}  Measurements of non-resonant $\chi^{(3)}$ of SWCNTs have been performed using some third-order nonlinear optical processes, including four-wave mixing\cite{KimetAl09NL,MyllyperkietAl10ACS} and nonlinear refraction and absorption,\cite{VivienetAl00OC,TatsuuraetAl03AM,ElimetAl04APL,RozhinetAl04TSF,SeoetAl06JPCS,WangetAl09JMC,MulleretAl10AO} where $\chi^{(3)}$ was measured to be on the order of $~10^{-10}$ to $10^{-12}$\,esu.
De Dominicis {\it et al}.\cite{DominicisetAl04APL}~observed third harmonic generation (THG) from SWCNT films using nanosecond pulses of 1064\,nm radiation, but the absolute value of $\chi^{(3)}$ was not quoted.

Here, we have made the determination of the value of the $\chi^{(3)}$ for SWCNTs via THG, by comparing our third harmonic intensity to the intensity of a reference material with well-known $\chi^{(3)}$. We have also made the determination of the strengths of all the nonzero tensor components of the $\chi^{(3)}$ tensor, by performing polarization dependence experiments on our highly aligned SWCNT sample.



We measured THG from highly-aligned SWCNT films were fabricated via the process described in Ref.~\onlinecite{PintetAl10ACS}. To make the sample, carbon nanotube arrays were grown vertically via chemical vapor deposition on a Fe catalyst-lined substrate.  The lines of catalysts were separated by a distance of 50\,$\mu$m, and the self-supporting carbon nanotube arrays were grown to a height specified by growth time.  The vertically aligned carpet was then separated from the catalyst substrate via post-growth H$_2$/H$_2$O vapor etch to release chemical bonds between catalyst particles and the
nanotubes and then deposited horizontally onto a sapphire substrate.  This process resulted in a large-area thin film of long, extremely well-aligned carbon nanotubes on sapphire, with a small amount of overlap.  The film thickness was measured via vertical scanning interferometry to be 1.6\,$\mu$m.  The films acted as nearly perfect polarizers for terahertz light,\cite{RenetAl09NL,RenetAl12NL,RenetAl13PRB} indicating the high degree of alignment of the SWCNTs.


THG measurements were performed using linearly-polarized midinfrared (MIR) radiation from an optical parametric amplifier pumped by a Ti:Sapphire-based chirped pulse amplifier (CPA-2010, Clark-MXR, Inc.).  The MIR pulses generated had a wavelength of 2.1\,$\mu$m, a repetition rate of 1\,kHz, a pulse-width of $\sim$300\,fs, and a pulse energy of $\sim$10\,$\mu$J.  The MIR beam was incident on a CaF$_2$ window that acted as a 97:3 beam splitter, where the reflected light was used to measure the fundamental intensity with a liquid nitrogen cooled mercury cadmium telluride detector.  The transmitted radiation was focused down to a spot size of $\sim$100\,$\mu$m in diameter, 
which allowed us to attain pump fluences of up to 100\,mJ/cm$^2$.  Our sample was placed at normal incidence at the focal position, and a third harmonic signal at 700\,nm was produced.  The third harmonic beam was filtered by a monochromator and measured by a photomultiplier tube.
%
%
In power-dependent experiments, the fundamental passed through a variable neutral density filter, to compare the measured third harmonic signal versus the incident fundamental power.  In polarization-dependent experiments, the third harmonic signal was passed through a rotatable polarizer such that the measured signal was either perpendicular or parallel to the incident fundamental polarization. The sample was then rotated through an angle $\phi$ about its normal, to determine the nonzero tensor components that contribute to the overall third harmonic signal.

\begin{figure}
\includegraphics[scale=.58]{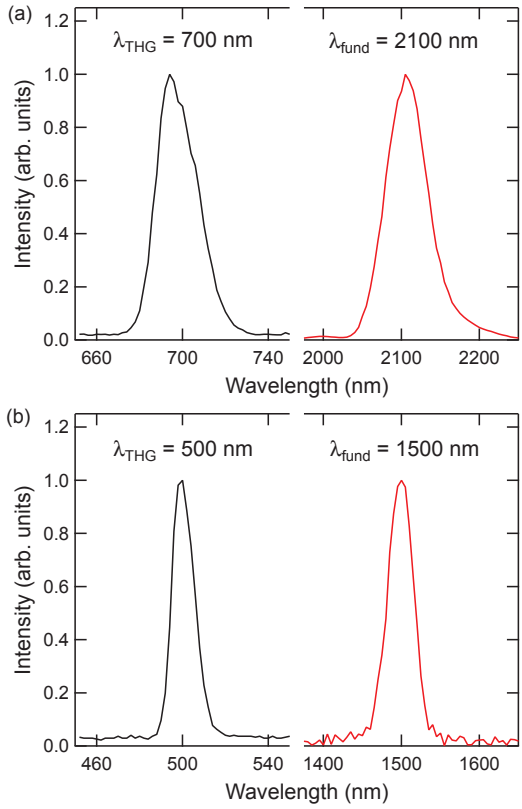}
\caption{(color online) (a)~Third harmonic spectrum generated at 700\,nm (left, black) from highly aligned SWCNTs on sapphire, and the fundamental spectrum (2100\,nm, right, red). (b)~Shift in third harmonic wavelength (left, black) from 700 to 500\,nm due to a shift in the fundamental from 2.1 to 1.5\,$\mu$m (right, red).  The SWCNTs are aligned parallel to the incident fundamental, and the induced third harmonic is polarized parallel to the fundamental.}
\label{Spectra}
\end{figure}

Figure~\ref{Spectra}(a) shows a spectrally resolved fundamental signal at 2.1\,$\mu$m and its third harmonic signal at 700\,nm generated from our SWCNT sample. Changing the fundamental input wavelength caused a subsequent shift in the third harmonic signal, as shown in Fig.~\ref{Spectra}(b). The bulk sapphire substrate did not produce a measurable third harmonic signal, and thus, the total measured signal can be attributed to the nanotube film.

\begin{figure}
\includegraphics[scale=.65]{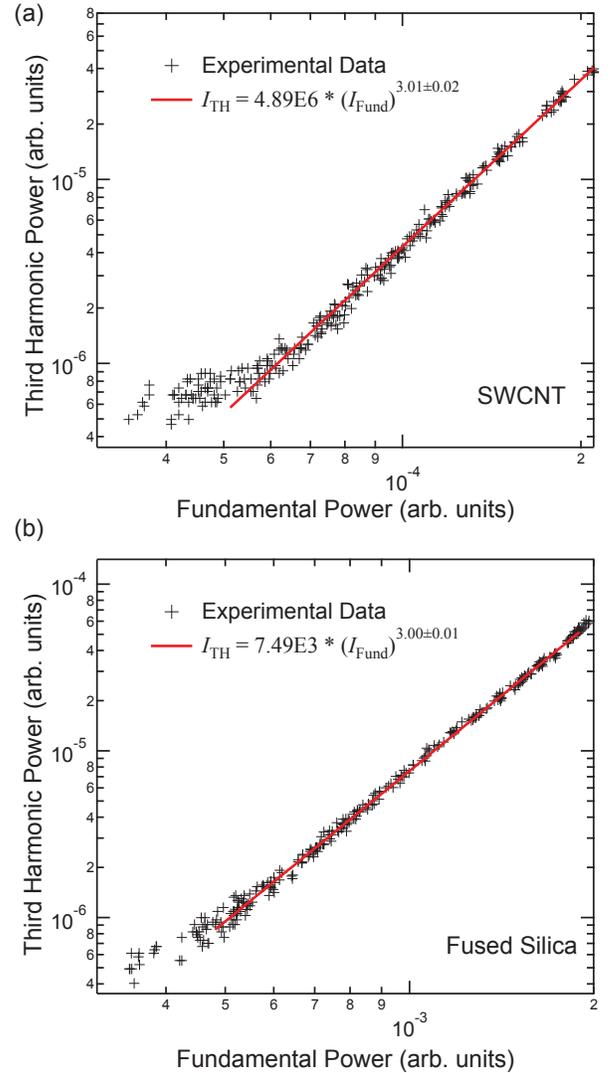}
\caption{(color online) Power dependence of the $700$\,nm third harmonic versus the 2100\,nm fundamental for the (a)~1.6\,$\mu$m-thick highly-aligned SWCNT sample and the (b)~26.5\,$\mu$m-thick fused silica reference.  Third harmonic shows cubic power dependence with the incident fundamental, as shown by the slope of the log/log plot (red).}
\label{PowerDep}
\end{figure}

Figure~\ref{PowerDep} shows that the intensity of the third harmonic produced by the SWCNT sample and fused silica reference varies with the cube of the fundamental, as shown by the straight line power fits on this log/log plot.  It can be seen that the same intensity produced by the carbon nanotube sample is achieved at an order of magnitude less input fundamental power.  This order of magnitude difference between the fused silica and the carbon nanotube cases shows that the $\chi^{(3)}$ for the SWCNT sample is about two orders of magnitude larger than that for fused silica, as suggested by the coefficients of the power fits.

In order to extract $\chi^{(3)}$ values for both samples from our experimental data, we need to consider phase matching conditions.  The intensity of the third harmonic generated in a film of thickness $L$ is given by
\begin{equation}
I_{3\omega}= \frac{576 \pi^4}{n_{3\omega} n^{3}_{\omega} \lambda^{2}_{\omega} c^2}\left| \chi^{(3)} \right|^2 I^{3}_{\omega} L^2 \frac{\sin^2(\Delta k L/2)}{(\Delta k L/2)^2},
\label{thginten}
\end{equation}
where $I_{\omega}$ ($\lambda_{\omega}$) is the intensity (wavelength) of the fundamental, $\Delta k \equiv k_{3\omega} - 3k_{\omega}$ is the phase mismatch between the third harmonic and the fundamental, and $n_{\omega}$ and $n_{3\omega}$ refer to the index of refraction at the fundamental and third harmonic frequencies, respectively.
The so-called coherence length $L_c = 2\pi / |\Delta k|  = \lambda_{\omega} / 3 \left|n_{3\omega}-n_\omega\right|$ is a measure of the distance over which the fundamental and third harmonic remain in phase. 
Because $\chi^{(3)}$ is a fourth-rank tensor, the $i^{\rm th}$ polarization component of the generated third harmonic field after traveling through a medium of length $L$ is defined as
\begin{equation}
E_{3\omega,i} = A_{\rm CNT} \left [\sum_{j,k,l} \chi_{ijkl}^{(3)} E_{\omega,j} E_{\omega,k} E_{\omega,l} \right ],
\label{2omaga-1omega}
\end{equation}
where the quantity in the bracket is the nonlinear dielectric polarization of the third harmonic, $E_{\omega, j}$, $E_{\omega, k}$, and $E_{\omega, l}$ are the fundamental field in the $j$, $k$, and $l$ directions, and $A_{\rm CNT}$ is a constant containing information about the fundamental-third harmonic interaction:
\begin{equation}
A_{\rm CNT} =\frac{24 \pi^2}{\sqrt{n_{3\omega} n^{3}_{\omega}} \lambda_{\omega} c} L \left|\frac{\sin(\Delta k L/2)}{\Delta k L/2}\right|e^{i\Delta kL}.
\label{Acnt}
\end{equation}

Because the SWCNT film is made of mostly air (as shown below), we assume the index of refraction to be nearly 1 for both the fundamental and third harmonic frequencies. Further, the sample thickness $L$ is quite small compared to the expected coherence length. Thus, we take the the phase mismatch for the third harmonic and fundamental to be negligible: $\Delta k L \ll \pi / 2$.

The thickness of the fused silica reference is 26.5\,$\mu$m (measured via Fabry-Perot fringes using Fourier transform infrared spectroscopy), which is within the calculated group-velocity walk-off length assuming 300$\,$fs pulses and on the order of the coherence length, $L_c$ = 17\,$\mu$m.  Because of the finite phase mismatch, the third harmonic intensity produced by the fused silica reference sample depends somewhat sensitively on the sample thickness. The measured intensity for our fused silica reference is less than the signal that would be produced if the sample thickness were exactly equal to the coherence length of 17$\,\mu$m. This must be taken into account since the quoted value of $\chi^{(3)}$ for fused silica was measured at a multiple of the coherence length.\cite{BuchalterMeredith82AO}

By taking the ratio of the third-harmonic intensities of the SWCNT sample and the fused silica reference and solving for $\chi^{(3)}_{\rm CNT}$, we obtain
\begin{eqnarray}
\chi^{(3)}_{\rm CNT}&=&\sqrt{n_{3\omega,\rm FS} n^{3}_{\omega,\rm FS}} \sqrt{\frac{I^{3}_{\omega,\rm FS}}{I^{3}_{\omega,\rm CNT}} }\chi^{(3)}_{\rm FS}\times \nonumber \\
 &&\sqrt{\frac{I_{3\omega,\rm CNT}}{I_{3\omega,\rm FS}}} \frac{L_{\rm CNT} }{L_{\rm FS}\left|\textrm{sinc} (\Delta k_{\rm FS} L_{\rm FS}/2) \right|}.
\label{comparison}
\end{eqnarray}
Because each of the power-dependence plots [Fig.~\ref{PowerDep}(a) and Fig.~\ref{PowerDep}(b)] can be fit with a power function,
\begin{equation}
I_{3\omega}=CI_{3\omega}^3,
\end{equation}
we compare the coefficients, $C$, to calculate the value of $\chi^{(3)}$ for the SWCNT film, using the relationship
\begin{equation}
\frac{C_{\rm CNT}}{C_{\rm FS}}=\frac{I_{3\omega,CNT}}{I_{3\omega,FS}}\frac{I_{3\omega,FS}^3}{I_{3\omega,CNT}^3}.
\end{equation}
By inserting all of our known quantities into Eq.~(\ref{comparison}), we find that the absolute value of $\chi^{(3)}$ for the SWCNT film is 5.53 $\times$ 10$^{-12}$\,esu.
%
However, it is important to note that the SWCNT film is not fully dense.  The carbon density of our film was measured\cite{PintetAl09APL} to be 60\,mg/cm$^3$, with which we can calculate the filling factor to be 8.75\%.  Taking into account this filling factor, we find the value of $\chi^{(3)}$ of a fully dense film to be 6.32 $\times$ 10$^{-11}$\,esu, which is extremely high as non-resonant $\chi^{(3)}$ for any material.

\begin{figure}
\begin{center}
\includegraphics[scale=.62]{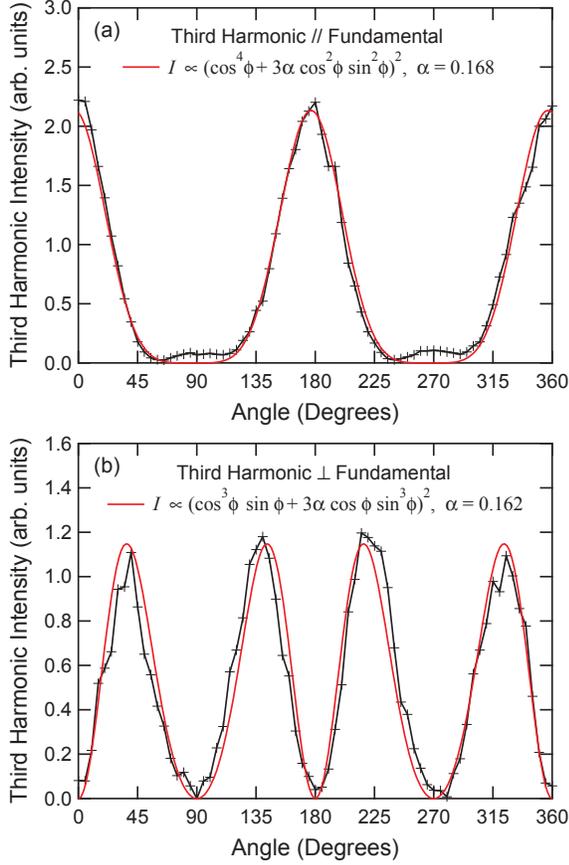}
\end{center}
\caption{(color online) Experimentally measured and theoretically calculated angular dependence for a THG signal polarized (a) parallel and (b) perpendicular to the fundamental, considering the $\chi^{(3)}$ tensor contribution relationship is $\alpha \chi^{(3)}_{zzzz}= \chi^{(3)}_{zzxx}$. The theoretical fits (red) show $\phi$ dependence for $\alpha \approx 1/6$. $\phi=0$ corresponds to light polarization parallel to the nanotube axis.}
\label{AngleDep}
\end{figure}

Far from resonances, one can make an order-of-magnitude estimate for $\chi^{(3)}$ as
\begin{equation}
\chi^{(3)} \sim \displaystyle  \chi^{(1)}(\omega) \left(\frac{\mu}{\hbar \omega} \right)^2,
\label{Alexey-eq1}
\end{equation}
where  $\mu = e \gamma/(E_c-E_v) \simeq e v_F/\omega$ is the dipole matrix element of the interband optical transition\cite{Goupalov05PRB2} between valence and conduction band states of energies $E_{v}$ and $E_{c}$, $\gamma$ = $(\sqrt{3}/2) a \gamma_0 \equiv \hbar v_F$, $a$ = 2.46~\AA, $\gamma_0$ = 2.89~eV is the transfer integral, $v_F \simeq c/300$ is the Fermi velocity of graphene. An order-of-magnitude estimate of $\chi^{(1)}$ in the effective-mass $kP$ description\cite{Ando05JPSJ} can be obtained by 
\begin{equation}
\chi^{(1)} \sim g \sum_{k,n} \frac{\mu_k^2 (f_v(k) - f_c(k))}{A(E_c(k) -E_v(k)  - \hbar \omega + i\delta)},
\label{Alexey-eq2}
\end{equation}
where summation is performed over all one-dimensional electron $k$-states, $A = \pi R_{\rm t}^2$, $R_{\rm t}$ is the nanotube radius, $g$ = 4 is the total degeneracy of an electron $k$-state, $E_{c,v}(k) = \gamma \sqrt{\kappa(n)^2 + k^2}$, $\kappa(n) = (1/R_{\rm t}) (n-\nu/3)$, and the quantum numbers $n$ and $\nu$  are defined in Ref.~\onlinecite{Ando05JPSJ}.  When far from any resonances and van Hove singularities at $k = 0$, one can replace the summation by a typical wave number involved in the optical transition, $k/(2\pi) \sim \hbar \omega/(4 \pi \gamma)$, and take $\mu \sim e v_F/\omega$. Assuming the difference in the occupation numbers $f_v-f_c$ to be equal to 1, taking the average radius $R$ = 3~nm,\cite{PintetAl10ACS} and using the frequency corresponding to the 2.1~$\mu$m wavelength, we arrive at $\chi^{(1)} \sim \mu^2/(\pi A \gamma)$ and $\chi^{(3)} \sim 9.8 \times 10^{-11}$~esu, in reasonable agreement with the measured value.

The magnitude of $\chi^{(3)}$ for SWCNTs can be also estimated from what is known for graphene.  The $\chi^{(3)}$ of graphene at near-infrared wavelengths has been measured to be $\sim$10$^{-7}$~esu,\cite{HandryetAl10PRL} which is the ``bulk'' susceptibility of a ``graphene material'' obtained by dividing the measured two-dimensional (sheet) susceptibility by the thickness, $d_{\rm g}$, of monolayer graphene.  Using Eq.~(\ref{Alexey-eq1}) with $\chi^{(1)} \sim e^2/(\hbar \omega d_{\rm g})$ and the value of $\omega$ corresponding to our pump wavelength (2.1~$\mu$m), we obtain $\chi^{(3)}$ $\sim$ 3 $\times$ 10$^{-8}$~esu, similar to the bulk susceptibility mentioned above.  The effective $\chi^{(3)}$ of a ``SWCNT material'' can then be obtained by ``rolling up'' graphene, i.e., diluting the volume by $R_{\rm t}^2/d_{\rm g}^2$ $\sim$ 150, which leads to the same ballpark estimate of $\chi^{(3)} \sim 10^{-10}$~esu. 

Since our SWCNT film contains macroscopically aligned ultralong nanotubes, we can also determine the components of the $\chi^{(3)}$ tensor.  The measured third harmonic signal had both parallel and perpendicular polarization components relative to the incident fundamental polarization, as shown in Figs.~\ref{AngleDep}(a) and \ref{AngleDep}(b), respectively.  There is strong $\phi$ dependence in the third harmonic signal both when the third harmonic is polarized parallel and perpendicular to the fundamental.  When the third harmonic is polarized parallel to the fundamental [Fig.~\ref{AngleDep}(a)], the intensity measured when the nanotube axis is parallel to the fundamental polarization ($\phi$ = 0$^{\circ}$ and 180$^{\circ}$) is almost two orders of magnitude larger
than that of the case where the nanotube axis is perpendicular to the fundamental ($\phi$ = 90$^{\circ}$ and 270$^{\circ}$).  When the measured third harmonic is polarized perpendicular to the fundamental, there are four peaks and valleys as the nanotube direction $\phi$ is scanned.  In both the parallel and perpendicular cases, the measured third harmonic is nearly zero when the fundamental polarization is perpendicular to the nanotube axis.

These polarization-dependent results give us significant insight into the relevant nonzero $\chi^{(3)}$ tensor elements for SWCNTs.  Here we use Eq.~(\ref{2omaga-1omega}) to analyze these data.  The incident fundamental is polarized parallel to the nanotube axis ($z$ axis) when $\phi=0^{\circ}$.  When the sample is rotated through an angle $\phi$ about its normal, we project this fundamental onto the axial ($z$) and radial ($x$) directions of the carbon nanotubes as
\begin{equation}
\vec{E}_\omega=E_\omega (\hat{e}_{z} \cos\phi - \hat{e}_{x} \sin\phi).
\end{equation}
By symmetry, the nonzero components of the $\chi^{(3)}$ tensor for a SWCNT are required to obey the following relationships\cite{LasseetAl03NL,Sundberg77JCP}
\begin{equation}
\alpha\chi_{zzzz}^{(3)}=\chi_{zzyy}^{(3)}=\chi_{zyzy}^{(3)}=
\chi_{zyyz}^{(3)}=\chi_{zzxx}^{(3)}=\chi_{zxzx}^{(3)}=\chi_{zxxz}^{(3)},
\end{equation}
where $\chi_{zzzz}^{(3)}\equiv \chi_{\rm CNT}^{(3)}$ and $\alpha $ ($0 < \alpha < 1$) is the ratio of the weaker tensor components to the dominant tensor component, $\chi_{zzzz}^{(3)}$.  Thus, the third harmonic field along the nanotube axis is
\begin{equation}
E_{3\omega,z} = A_{\rm CNT}\chi_{\rm CNT}^{(3)}E^3_\omega (\cos^3\phi+3\alpha\cos\phi\sin^2\phi).
\end{equation}
After projecting the nanotube coordinate system back to the coordinate system of the incident fundamental, we find that the component of the induced third harmonic field parallel to the incident fundamental is
\begin{equation}
E_{3\omega,\parallel} = A_{\rm CNT}\chi_{\rm CNT}^{(3)}E^3_\omega (\cos^3\phi+3\alpha \cos\phi\sin^2\phi)\cos\phi,
\end{equation}
and the corresponding intensity is
\begin{equation}
I_{3\omega,\parallel} = |A_{\rm CNT}|^2|\chi_{\rm CNT}^{(3)}|^2 I^3_\omega (\cos^4\phi+3\alpha\cos^2\phi\sin^2\phi)^2.
\label{parallelfit}
\end{equation}
Simulations show that the peak that would arise purely from the dominant tensor component $\chi_{zzzz}^{(3)}$ becomes broadened as the contribution from the weaker tensor components increases (i.e., as $\alpha$ increases from zero).  Similarly, the induced third harmonic intensity polarized perpendicular to the incident fundamental is
\begin{equation}
I_{3\omega,\bot} = |A_{\rm CNT}|^2|\chi_{\rm CNT}^{(3)}|^2 I^3_\omega (\cos^3\phi\sin\phi+3\alpha\cos\phi\sin^3\phi)^2.
\label{perpfit}
\end{equation}
As the contributions from the weaker tensor components increase (i.e., as $\alpha$ increases from zero), the peaks of the induced third harmonic shift and sharpen.

We fit our data with the theoretically determined fitting functions in Eqs.~(\ref{parallelfit}) and (\ref{perpfit}). By allowing the value of $\alpha$ to be the only adjustable parameter, we found that the measured third harmonic signals (black line with cross markers) are in excellent agreement with the theoretically calculated fits (red line), as shown in Figs.~\ref{AngleDep}(a) and \ref{AngleDep}(b). Not only do the fits correlate extremely well with the measurements, but the fit parameter $\alpha$ is approximately equal to 1/6 in both cases. This value indicates that the dominant $\chi^{(3)}$ tensor component, $\chi^{(3)}_{zzzz}$, is six times larger than the weaker components.

In Fig.~\ref{AngleDep}(a), the third harmonic signal is small but finite when the SWCNTs are oriented perpendicular to the incident fundamental ($\phi$ = 90$^{\circ}$ and 270$^{\circ}$), whereas theory based on either intraband or interband transitions predicts zero third harmonic signal at these points.  Contributions from cross-polarized transitions, e.g., $E_n$ to $E_{n \pm 1}$, could be a source of this finite signal in the perpendicular configuration although such transitions are weakened by the depolarization effect.\cite{AjikiAndo94Physica}


In summary, we successfully observed third harmonic generation in highly aligned SWCNTs on a sapphire substrate. Through power-dependent measurements,
we were able to determine $\chi^{(3)}$ to be 5.53 $\times$ 10$^{-12}$~esu for the film.  With an estimate filling factor of 8.75\% for our film, $\chi^{(3)}$ value of a fully dense film would be 6.32 $\times$ 10$^{-11}$~esu.  Furthermore, through orientation-dependent experiments we were successfully able to extract all the relevant nonzero $\chi^{(3)}$ tensor elements.  We found that weaker $\chi^{(3)}$ tensor elements are approximately 1/6 the strength of the dominant $\chi^{(3)}_{zzzz}$.

\smallskip

We acknowledge support from the National Science Foundation (through Grant No.~EEC-0540832), Department of Energy BES Program (through Grant No.~DE-FG02-06ER46308), and the Robert A.~Welch Foundation (through Grant No.~C-1509).  
We thank Deleon J.~L.~Reescano for his assistance.




\end{document}